\documentclass[runningheads]{llncs}

\usepackage{amsmath,amssymb}	
\usepackage{svg}
\usepackage{graphicx}

\newcommand\blfootnote[1]{%
  \begingroup
  \renewcommand\thefootnote{}\footnote{#1}%
  \addtocounter{footnote}{-1}%
  \endgroup
}

\def\Plus{\texttt{+}}
\usepackage[hyperfootnotes=false]{hyperref}
\usepackage{subcaption} 

\begin{document}

\title{Unsupervised 3D Brain Anomaly Detection}

\author{Jaime Simarro Viana \inst{1,2} \and
Ezequiel de la Rosa\inst{1,3} \and Thijs Vande Vyvere \inst{1,4} \and
David Robben \inst{1,5,6} \and
Diana M. Sima\inst{1};
CENTER-TBI Participants and Investigators \thanks{CENTER-TBI participants and investigators are listed at the end of the supplementary material.} }

\authorrunning{J. Simarro et al.}

\institute{ico\textbf{metrix}, Research and Development, Leuven, Belgium.
\and 
Erasmus Joint Master In Medical Imaging and Applications, University of Girona, Girona, Spain
\and
Department of Computer Science, Technical University of Munich, Munich, Germany
\and 
Department of Radiology, Neuroradiology Division, Antwerp University Hospital and University of Antwerp, Antwerp, Belgium
\and
Medical Image Computing (MIC), ESAT-PSI, Department of Electrical Engineering, KU Leuven, Leuven, Belgium
\and
Medical Imaging Research Center (MIRC), KU Leuven, Leuven, Belgium
}
\maketitle       

\begin{abstract}

Anomaly detection (AD) is the identification of data samples that do not fit a learned data distribution. As such, AD systems can help physicians to determine the presence, severity, and extension of a pathology. Deep generative models, such as Generative Adversarial Networks (GANs), can be exploited to capture anatomical variability. Consequently, any outlier (i.e., sample falling outside of the learned distribution) can be detected as an abnormality in an unsupervised fashion. By using this method, we can not only detect expected or known lesions, but we can even unveil previously unrecognized biomarkers. To the best of our knowledge, this study exemplifies the first AD approach that can efficiently handle volumetric data and detect 3D brain anomalies in one single model. Our proposal is a volumetric and high-detail extension of the 2D f-AnoGAN model obtained by combining a state-of-the-art 3D GAN with refinement training steps. In experiments using non-contrast computed tomography images from traumatic brain injury (TBI) patients, the model detects and localizes TBI abnormalities with an area under the ROC curve of $\sim$75$\%$. Moreover, we test the potential of the method for detecting other anomalies such as low quality images, preprocessing inaccuracies, artifacts, and even the presence of post-operative signs (such as a craniectomy or a brain shunt). The method has potential for rapidly labeling abnormalities in massive imaging datasets, as well as identifying new biomarkers.

 \blfootnote{\\ \textit{The final authenticated publication is available online at \url{https://doi.org/10.1007/978-3-030-72084-1_13}}}
\end{abstract}

\section{Introduction}
\blfootnote{\\ \textit{The final authenticated publication is available online at \url{https://doi.org/10.1007/978-3-030-72084-1_13}}}
\label{sec:introduction}
\textit{Supervised} deep learning techniques have shown outstanding performance in a wide diversity of medical imaging tasks, and can even outperform radiologists in areas such as lung cancer detection \cite{lungCancer} or breast tumor identification \cite{breastCancer}. However, these techniques require large annotated databases, which are expensive and time-consuming to obtain \cite{unsupervisedOCT}. Furthermore, manual annotations often are disease-specific and do not always cover the wide range of abnormalities that can be present in a scan \cite{siamese,uncertainty}. In contrast, \textit{unsupervised} learning models are capable of discovering patterns from label-free databases. A current challenge in this field is unsupervised \textit{anomaly detection (AD)}. AD is the task of identifying test data that does not fit the data distribution seen during training \cite{AnoGAN}. In clinical practice, AD represents a crucial step. Physicians learn the normal anatomical variability and they recognize anomalies by implicitly comparing to normal cases or healthy surrounding areas. As such, many AD models identify abnormalities in an unconstrained fashion by mimicking this human behavior.

\textbf{State of the art.}
Deep generative models, such as Variational Auto-encoders (VAEs) \cite{VAE} and Generative Adversarial Networks (GANs) \cite{GAN}, are able to generate synthetic images that capture the variability of the training images. Thus, if a deep generative model is trained over lesion-free data, anomalies could be discovered by detecting samples that do not fit this lesion-free variability. For AD in retina images, Schlegl et al. \cite{AnoGAN} suggest that a GAN trained on healthy images should not be able to reconstruct abnormalities. In that work, a slow iterative optimization algorithm is used to find the GAN's latent space projection of a given image. To make this mapping technique faster, Schlegl et al. \cite{fAnoGAN} propose f-AnoGAN, which replaces the iterative algorithm with an encoder network. In brain imaging, most recent AD work has focused on 2D axial images. Baur et al. \cite{AnoVAEGAN} use a combination of a spatial VAE and an adversarial network for delineating multiple sclerosis lesions in MR images. You et al. \cite{VAErestoration} detect brain tumors using a Gaussian Mixture VAE with restoration of the latent space, while Pawlowski et al. \cite{bayesianAE} use Bayesian Auto-encoders to detect traumatic brain injury lesions. In a very recent comparative study on brain AD, the performance of f-AnoGAN is remarkable in diverse datasets \cite{ComparativeStudy}. All of these 2D-based approaches have several drawbacks: i) they do not consider volumetric information and, consequently, they do not effectively handle the complex brain anatomy; ii) they have to consider the whole brain image since there is no prior information of the anomaly localization; iii) they require multiple models for evaluating an entire scan.

\textbf{Contributions of this work.}
We propose, to the best of our knowledge, the first 3D brain anomaly detector. This model effectively handles complex brain structures and provides reliable 3D reconstructions based on brain anatomy. The present work is inspired by the 2D f-AnoGAN architecture. However, the proposed methodology differs in several aspects from f-AnoGAN: i) the network learns from \textit{volumetric information} creating 3D image reconstructions; ii) the architecture is enhanced by using a modified version of a state-of-the-art 3D GAN; iii) a new training step is proposed to deal with the lack of details in reconstruction images. We show the AD capability of our proposed method in two independent traumatic brain injury (TBI) datasets. Besides, we evaluate its potential for AD in postsurgical cases and poor quality scans.

\section{Methods}
\label{sec:methods}
\textbf{Database.}
For devising and validating the approach, we use non-contrast computed tomography (NCCT) data of traumatic brain injury patients. TBI includes a vast spectrum of pathoanatomical anomalies that may affect any brain region. Two independent datasets are used for our experiments:

\begin{description}
  \item[$\ast$ CENTER-TBI.] The collaborative European NeuroTrauma Effectiveness Research in Traumatic Brain Injury (CENTER-TBI) project include a database collection of NCCT images \cite{center-TBI}. The study protocol was approved by the national and local ethics committees for each participating center. Informed consent, including use of data for other research purposes, was obtained in each subject according to local regulations. Patient data was de-identified and coded by means of a Global Unique Patient Identifier. In this multi-center, multi-scanner, longitudinal study, all the NCCT images of TBI patients were visually reviewed and the abnormal findings were reported in a structured way by an expert panel. We retrieve a selection of images from a centralized imaging repository that stores the data collected and sent by the different sites. This dataset includes brain images without NCCT abnormal findings by expert review ($n= 637$ total scans) and manually annotated TBI scans ($n = 102$) with abnormal NCCT findings.
  \item[$\ast$ PhysioNet.] The model is also tested on the publicly available database, online at the PhysioNet repository \footnote{ \url{https://physionet.org/content/ct-ich/1.3.1/}}\blfootnote{\textit{\\The final authenticated publication is available online at \url{https://doi.org/10.1007/978-3-030-72084-1_13}}} \cite{PhysioNet,PhysioNet131,ICH_PhysioNet}. This dataset includes 37 subjects without NCCT abnormal findings and 33 TBI patients.
\end{description}

The training of the model is performed over $\sim 80\%$ ($n = 532$) of the CENTER-TBI data without abnormal NCCT findings. As test sets we use CENTER-TBI (remaining $20\%$, $n = 105$ and all TBI cases with abnormal findings) and PhysioNet (entire database). 

\textbf{Preprocessing.} All scans undergo the following preprocessing steps:\begin{enumerate}
\item NCCT images are registered to the MNI space with an affine transformation. 
\item An automatic quality control process is performed using the FDA approved ico\textbf{brain} TBI software \cite{icobrain_tbi}. Highly corrupted images are automatically discarded.
\item Using the same software, a skull-stripping operation is performed.
\item Black boundaries caused by the application of the brain mask are removed.
\item After a Gaussian smoothing, images are resized to $64 \times 64 \times 64$ using linear interpolation.
\item A soft tissue windowing [-20, 100 HU] is performed, similarly as in \cite{preprocessingTBI}.
\item The images are globally min-max normalized between -1 and 1.
\end{enumerate}

\begin{figure}
 \centering
 \begin{minipage}[c]{0.44\textwidth}
  \includegraphics[width=1.1\linewidth]{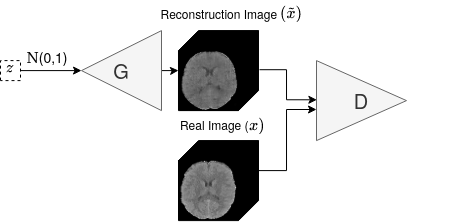}
\caption{3D GAN networks architecture.}
\label{fig:arch_GAN}
 \end{minipage}
 \hfill
 \begin{minipage}[c]{0.54\textwidth}
\begin{equation}
\min_G \max_D\mathbb{E}_{x\sim \mathbb{P}_{r}}[\log D(x)]+ \mathbb{E}_{\tilde{x}\sim \mathbb{P}_g}
[\log(1 - D(\tilde{x}))]
\label{eq:loss_GAN}
\end{equation}
\caption*{Eq. 1: Original GAN loss function. Where $\mathbb{P}_{r}$ is the real data distribution and $\mathbb{P}_{g}$ is the model distribution defined by $\tilde{x}=G(z)$. The input $z$ of $G$ is sampled from a Gaussian distribution,  ${N} (\mu=0,\sigma^{2}=1)$.}
 \end{minipage}
\end{figure}

\textbf{Training strategy.}  \blfootnote{\\ \textit{The final authenticated publication is available online at \url{https://doi.org/10.1007/978-3-030-72084-1_13}}}
As proposed in f-AnoGAN, the model framework is composed of a GAN and an encoder network. These networks are trained in a multi-step training strategy where brain images without NCCT abnormal findings are used. Then, these trained models are able to detect anomalies using an anomaly score. The training strategy is divided into the following training steps:

\textit{1. GAN training.}
The GAN (Fig. \ref{fig:arch_GAN}) training is based on a competitive game between two networks: the \textit{generator} network ($G$) and the \textit{discriminator} network ($D$). During training, $G$ maximizes the probability of $D$ making a mistake, while $D$ maximizes the probability of correctly predicting the real and generated samples. Eq. \ref{eq:loss_GAN} shows the objective function for parametrizing the model. 

The main challenge in 3D generation is the mode collapse problem \cite{alphaGan}. Thus, we use Wasserstein-1 distance (also called Earth-Mover) \cite{WGAN} in our GAN loss. Moreover, the gradient penalty \cite{gradientWGAN} is also included in order to increase training stability. The resulting discriminator and generator loss functions (respectively $L_{D}$ and $L_{G}$) are hence as follows: 

\begin{align} 
L_{D} &= \mathbb{E}_{\tilde{x}\sim \mathbb{P}_g}
[ D(\tilde{x})] -\mathbb{E}_{x\sim \mathbb{P}_{r}}[D(x)] +\lambda \mathbb{E}_{\hat{x}\sim \mathbb{P}_{\hat{x}}}[(\left \| \bigtriangledown_{\hat{x}}D(\hat{x}) \right \|_2-1)^{2}] \\ 
L_{G} &= -\mathbb{E}_{\tilde{x}\sim \mathbb{P}_g}[ D(\tilde{x})]
\end{align}
where $\mathbb{P}_{\hat{x}}$ is sampled uniformly along straight lines between a pair of points sampled from $\mathbb{P}_{r}$ and $\mathbb{P}_{g}$ and $\lambda$ is a weighting parameter.

\begin{figure}
 \centering
 \begin{minipage}[c]{0.49\textwidth}
  \includegraphics[width=1.1\linewidth]{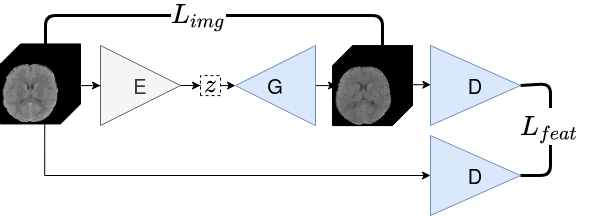}
\caption{Image space loss and discriminator feature space loss. Networks in blue do not change their weights during encoder training phase.}
\label{fig:enc_traini_2error}
 \end{minipage}
 \hfill
 \begin{minipage}[c]{0.49\textwidth}
\begin{equation}
L_{img}=\frac{1}{n}\left \|x-\tilde{x} \right \|^{2}
\label{eq:image_loss}
\end{equation}
\begin{equation}
  L_{feat}=\frac{1}{m}\left\|f(x)-f(\tilde{x}) \right \|^{2}
  \label{eq:feature_loss}
\end{equation}
\begin{equation}
  L_{E}=L_{img}+\kappa \cdot L_{feat}
  \label{eq:enc_loss}
\end{equation}
\caption*{Eq. [4-6]: $n$ is the number of voxels in the image, $m$ is the dimensionality of the discriminator feature space, $f(x)$ the activation on the intermediate layer of $D$ and $\kappa$ is a weighting parameter.}
 \end{minipage}
\end{figure}

\textit{2. Encoder training.}
Once the adversarial training is completed, $G$ knows how to map from the latent space ($z$) to an image ($\tilde{x}$), $G(z)=z\rightarrow \tilde{x}$. However, the representation of a given image in the latent space is unknown. The \textit{encoder} network ($E$) makes this mapping, $E(x)=x\rightarrow z$. As shown in Fig.~ \ref{fig:enc_traini_2error}, the weights of $G$ and $D$ remain frozen while only the weights of $E$ are optimized. Results show that the $E$ network exploits proper latent space representations and, therefore, $G$ outcomes good reconstructions without requiring a forced constraint over $z$. 
The $E$ network is optimized by minimizing $L_{E}$, a weighted sum of: \textit{image space loss ($L_{img}$)} and \textit{discriminator feature space loss ($L_{feat}$)} (see Eq. \ref{eq:image_loss}-\ref{eq:enc_loss}). The use of this feature space is suggested by \cite{AnoGAN} and is inspired by the feature matching technique \cite{featureMatching}.

\textit{3. Techniques for improving the performance.}\blfootnote{\\ \textit{The final authenticated publication is available online at \url{https://doi.org/10.1007/978-3-030-72084-1_13}}}
After preliminary experiments, a lack of details in the reconstructed images is observed. Therefore, we propose a new learning step that provides explicit learning feedback of the vast information that a 3D image contains. This extra training step provides a fine-tuning of the networks weights rather than a full model training from scratch. Empirical results show that optimizing the weights of $E$ and $G$ while minimizing $L_{Encoder}$ is the most convenient training strategy (see supplementary material Table S2).

\textbf{Anomaly score.}
The anomaly score quantifies the deviation of test images and corresponding reconstruction \cite{fAnoGAN}. Note that the reconstructions are generated by considering only the distribution of the data used for training, i.e., NCCT images without any radiological findings. Therefore, this anomaly score can be interpreted as a distance metric between the input image and the learned anatomical variability. As presented in f-AnoGAN, the anomaly score for a given image is obtained using the function shown in Eq. \ref{eq:enc_loss}. Thresholding the anomaly score provides a global classification of an image as abnormal or not. AD performance can thus be evaluated using a ROC analysis of the anomaly score.

\begin{figure*}[h!]
  \centering
  \includegraphics[width=1\textwidth]{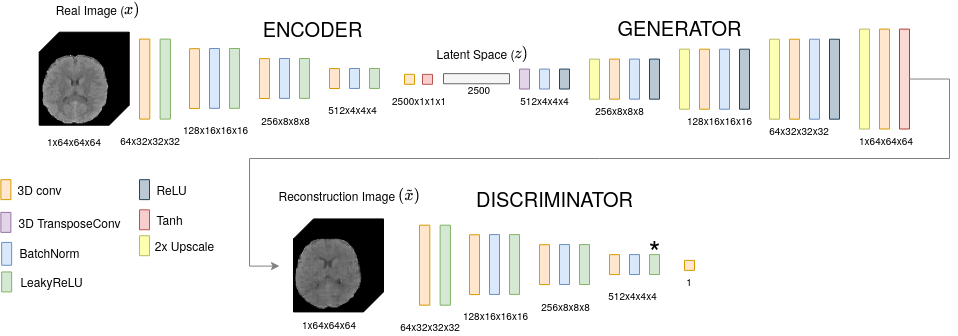}
  \caption{Architecture details. Dimensions of each feature maps are shown for each block. The asterisk (*) denotes the intermediate layer of the $D$ used in $f(x)$.}
  \label{fig:architecture_details}
\end{figure*}

\textbf{Model architecture.} A state-of-the-art brain 3D GAN architecture \cite{alphaGan} is used as foundation for the AD model. We add a hyperbolic tangent activation in the last encoder. We also increase the original latent space dimension to 2500. Architecture details are shown in Fig. \ref{fig:architecture_details}.

\section{Results and Discussion} \blfootnote{\\ \textit{The final authenticated publication is available online at \url{https://doi.org/10.1007/978-3-030-72084-1_13}}}
\label{sec:results}

\textbf{Comparison with 2D f-AnoGAN.}
Axial slices within a middle brain range (ensuring similar anatomical structures) are randomly selected to train a 2D model. In order to have a fair comparison in terms of AD of TBI abnormalities, images with abnormalities that are not located in the selected slice are discarded. In our experiments, the 3D model outperforms the 2D one by 4$\%$ in the area under the ROC curve.

\begin{figure}
 \centering
 \begin{minipage}[b]{0.49\textwidth}
 \includegraphics[width=1\linewidth]{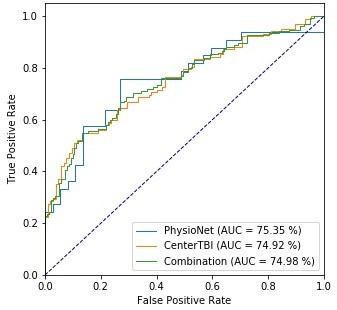}
\caption{Comparison of ROC curves for the different datasets. AUC: Area under the ROC curve.}
\label{fig:ROC}
 \end{minipage}
 \hfill
 \begin{minipage}[b]{0.49\textwidth}
\includegraphics[width=1\linewidth]{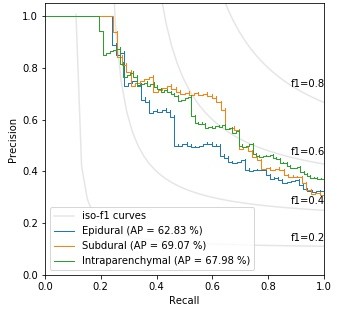}
 \caption{ Comparison of Precision Recall curve for each TBI lesion. AP: Average precision.}
  \label{fig:Presion-Recall_lesions}
 \end{minipage}
\end{figure}

\textbf{AD performance.} 
In Fig. \ref{fig:ROC} we show a comparison of ROC curves for the different TBI datasets. The AD performance reaches $\sim 75\%$ area under the ROC in both databases. At the Youden index of the ROC curve over the combined datasets, the model has a $70.75\%$ of accuracy, $54.07\%$ of recall, and $86.66\%$ of specificity. Another operating point could be chosen, depending on the clinically desired balance between sensitivity and specificity.    

\textbf{AD performance by lesion type.}
Subjects from both datasets with at least one of the following hematomas are used to evaluate the performance of the model: \textit{epidural}, \textit{subdural}, and \textit{intraparenchymal}. If the model detects an anomalous case having one of these lesions, this is counted as a detection, no matter if other lesions are also present. The model performance is similar across lesion types, performing slightly better for subdural hematomas (see Fig. \ref{fig:Presion-Recall_lesions}).

\begin{figure*}[!ht]
 \centering
 \includegraphics[width=0.85\textwidth]{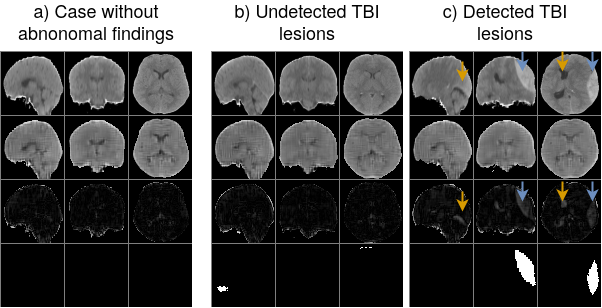}
 \caption{Anomalous region localization using voxel-level error. First row: Original image after preprocessing 
 ($x$). Second row: Reconstruction image ($\tilde{x}$). Third row: voxel-level error image. Fourth row: ground truth lesion segmentation. Arrows indicate the anomalies detected by the model; the blue ones show a labeled anomaly in the database while orange ones show an unlabeled anomaly.}
 \label{fig:localization}
\end{figure*}

\textbf{Qualitative results and anomaly localization.}\blfootnote{\\ \textit{The final authenticated publication is available online at \url{https://doi.org/10.1007/978-3-030-72084-1_13}}}
The method is able to localize anomalies through a voxel-wise subtraction of the original image $x$ and its reconstruction $\tilde{x}$. Fig. \ref{fig:localization} exemplifies the most common cases in anomaly localization. \textit{i) Case without abnormal findings:} In the second row of the figure, we can appreciate reliable reconstructions of the input images. No relevant region can be considered as a lesion in the voxel-wise error image (third row). \textit{ii) Undetected TBI lesions:} Tiny lesions can be missed inside the anatomical variability, so the reconstruction image matches with the original one, making the TBI lesion hardly detectable. \textit{iii) Detected TBI lesions:} If lesions fall outside the learned distribution, the model will not be able to reconstruct this region and it will select the closest representation that has been learned, which could be thought of as a \textit{healthy} brain representation. Hence, lesions are well localized (see blue arrows). Refer to supplementary material Fig. S1-S3 to visualize examples in different anatomical planes.

\textbf{AD for biomarker discovery.}
The proposed unsupervised learning model is capable of detecting unknown/unlabeled abnormalities. Fig. \ref{fig:localization}-c shows an epidural hematoma that has been labeled. It can be noticed that the lesion introduces a \textit{mass effect} affecting nearby structures: the lateral ventricles are compressed and displaced, and a \textit{midline shift} can be observed. The mass effect is not labeled in the dataset and, hence, no supervised approach would detect it. However, the proposed method overcomes this limitation, highlighting and locating this anomaly (see orange arrows). This property of detecting unlabeled anomalies can be used for \textit{biomarker discovery}.

\begin{figure}[h]
 \centering
 \includegraphics[width=1\textwidth]{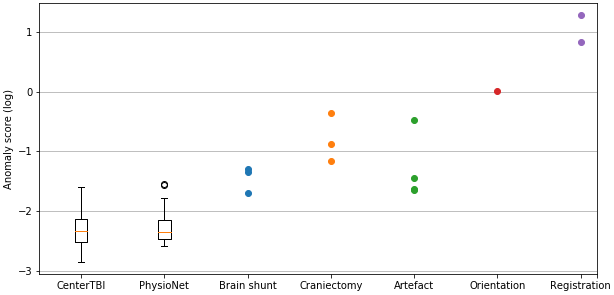}
 \caption{Different NCCT images: images without abnormal findings from CENTER-TBI and PhysioNet, followed by various post-surgical pathologies and low quality images. Logarithm of the anomaly scores are used for better visualization.}
 \label{fig:QC_boxplot}
\end{figure}

\textbf{AD for quality control.}
Given that an anomaly is defined as any type of data unrepresented by the normal data distribution, we can extend our AD model to detect any kind of \textit{outlier} sample. We evaluate its potential for detecting low quality images (such as artifacts, wrong registrations, and wrong orientations) and post-surgical signs (such as a craniectomy or brain shunt). Fig. \ref{fig:QC_boxplot} shows the results of this proof of concept application. Images with anomalies have much higher anomaly scores than the distribution without any radiological findings. 

\section{Conclusion}\blfootnote{\\ \textit{The final authenticated publication is available online at \url{https://doi.org/10.1007/978-3-030-72084-1_13}}}
\label{sec:Conclusion}

The present model is, to our knowledge, the first feasible attempt for 3D brain AD screening in a real-world scenario. We overcome several limitations of previous approaches: i) our model handles volumetric information; ii) it is capable of detecting a wide variety of abnormalities in an unconstrained manner; iii) in contrast with supervised learning techniques, the model is not biased towards expert annotations; iv) the model offers good generalization capabilities, providing database-invariant anomaly scores; v) the voxel-wise error image localizes abnormalities, increasing the model interpretability.

In our experiments, the GPU memory limited the input data resolution. As future perspectives, we consider working with higher resolution images ($>64 \times 64 \times 64$), which would help to detect small anomalies and, hence, improve the model performance. This improvement will reduce the difference between unsupervised TBI detection performance and supervised learning models such as \cite{supervised_deep,supervised_expert}.
Besides, the voxel-error image could be extended to perform 3D anomaly segmentation (i.e., generating anomaly masks). Also, we want to extend our model to work with MR images, which is much more challenging than working with NCCT. In MR, the trained GAN should capture both the anatomical brain variability and the intrinsic MR scans variability. In addition, the model could be improved by taking into consideration demographic variables such as age.

\section{Acknowledgments}\blfootnote{\\ \textit{The final authenticated publication is available online at \url{https://doi.org/10.1007/978-3-030-72084-1_13}}}

JSV received an Erasmus$\Plus$ scholarship from the Erasmus Mundus Joint Master Degree in Medical Imaging and Applications (MAIA), a programme funded by the Erasmus$\Plus$ programme of the European Union (EU grant 20152491). This project received funding from the European Union's Horizon 2020 research and innovation program under the Marie Sklodowska-Curie grant agreement TRABIT No 765148. DR is supported by an innovation mandate of Flanders Innovation \& Entrepreneurship (VLAIO). Data used in preparation of this manuscript were obtained in the context of CENTER-TBI, a large collaborative project with the support of the European Union 7th Framework program (EC grant 602150). Additional funding was obtained from the Hannelore Kohl Stiftung (Germany), from OneMind (USA) and from Integra LifeSciences Corporation (USA)

We thank Charlotte Timmermans and Nathan Vanalken for performing the manual TBI segmentations.

%
%

\bibliographystyle{splncs04}
\bibliography{brainLes}

\end{document}